# Extent of hydrogen coverage of Si(001) under chemical vapor deposition conditions from *ab initio* approaches

Phil Rosenow, Ralf Tonner[a]

*Fachbereich Chemie and Wissenschaftliches Zentrum für Materialwissenschaften, Philipps-Universität Marburg, Hans-Meerwein-Straße, 35032 Marburg, Germany*

The extent of hydrogen coverage of the Si(001)*c*(4x2) surface in the presence of hydrogen gas has been studied with dispersion corrected density functional theory. Electronic energy contributions are well described using a hybrid functional. The temperature dependence of the coverage in thermodynamic equilibrium was studied computing the phonon spectrum in a supercell approach. As an approximation to these demanding computations, an interpolated phonon approach was found to give comparable accuracy. The simpler *ab initio* thermodynamic approach is not accurate enough for the system studied, even if corrections by the Einstein model for surface vibrations are considered. The on-set of $H_2$ desorption from the fully hydrogenated surface is predicted to occur at temperatures around 750 K. Strong changes in hydrogen coverage are found between 1000 and 1200 K in good agreement with previous reflectance anisotropy spectroscopy experiments. These findings allow a rational choice for the surface state in the computational treatment of chemical reactions under typical metal organic vapor phase epitaxy conditions on Si(001).

---

[a] Author to whom correspondance should be adressed. Electronic mail: tonner@chemie.uni-marburg.de.

# I. INTRODUCTION

The functionalization of silicon surfaces with organic or inorganic molecules extends the range of applications for this important semiconductor material. One promising field is "silicon photonics", i.e. the construction of optically active devices based on silicon substrates.[1] The most important aim in this field is building a laser, for example by monolithically integrating an optically active layer on silicon substrate. Promising candidates here are quantum well structures with group 13 and group 15 elements (aka. III/V), for example the material system Ga(N, As, P).[2] These materials are usually produced by metal organic vapor phase epitaxy (MOVPE), a method that deposits liquid-state precursor molecules on heated surfaces via chemical vapor deposition.[3] The structure of the final thin-film and thus the properties of the resulting device are crucially determined by the chemical reactions in the nucleation phase of the deposition process.[4] This guides our ongoing research endeavors toward the chemical reactivity in the MOVPE process by quantum chemical methods.[5] Our intention to extend these investigations to surface reactivity requires an in-depth knowledge about the state of the surface under MOVPE conditions, which is the topic of this study.

The semiconductor surface most often used in MOVPE experiments is Si(001) due to its technological importance.[3] This surface undergoes the well-known dimer reconstruction and buckling leading to Si(001)$c$(4x2) being the pristine state (Scheme 1f).[6] The silicon atoms at the surface exhibit "dangling bonds", i.e. unpaired electrons which explains the high reactivity towards adsorbates.[7] Therefore, it is very likely that the commonly used carrier gas $H_2$ will react with the surface atoms under the elevated temperatures used for the deposition process. The final state for hydrogenation under most experimental conditions is one hydrogen atom per surface dimer atom (Scheme 1a), which is defined as one monolayer coverage ($\theta$ = 1 ML) is this study. The monohydride termination of the Si(001) surface under MOVPE conditions was confirmed with infrared spectroscopy.[8] The temperature dependence of this hydrogen coverage is an important factor for correct modelling of the growth reactions and is not well documented up to now. Experimental surface science studies are mostly performed under ultra-high vacuum (UHV) conditions and have given valuable insight in the dynamics of hydrogen adsorption, desorption and diffusion on the surface in the past.[9,10,11] Likewise, computational studies have been devoted to desorption mechanisms and surface dynamics.[12,13,14] Two studies using temperature programmed desorption (TPD) in UHV conditions show an onset of $H_2$-desorption from hydrogenated Si(001) surfaces to occur between 700-780 K[15] and 750±25 K[16], respectively. Brückner et al. studied the desorption of $H_2$ from hydrogenated Si(001) both in a flow of hydrogen and nitrogen carrier gas with reflectance anisotropy spectroscopy (RAS) under MOVPE conditions.[17] In a nitrogen flow, hydrogen desorbed from the H/Si(001) surface (Scheme 1a) at 750 K within 30 minutes leaving a pristine Si(001) surface in accordance with the TPD results cited above. Under typical MOVPE conditions with hydrogen carrier gas, dehydrogenation gradually increased with increasing temperatures. Lowering the temperature from 1270 K to 670 K, a strong change in the RAS signal was observed around T = 1070 K, which is associated with significant hydrogenation of the bare surface. At lower temperatures, the signal of the fully hydrogenated surface is found when compensating for a thermal drift in the measurements. The authors thus concluded that the silicon surface is hydrogenated for temperatures below 1070 K in a $H_2$-atmosphere and pristine at 1270 K. Extrapolation from the low pressures used in the study to atmospheric pressure of hydrogen yields $\theta$ = 0.94 ML coverage at 1070 K and $\theta$ = 0.25 ML coverage at 1270 K. It should be stressed, that we will focus solely on surface coverage in thermodynamic equilibrium – adsorption and desorption dynamics are not aimed at here and were extensively reviewed elsewhere.[9-14] Likewise, the H$_2$ induced formation of reconstruction domains[18] is not the aim of this study.

The theoretical treatment of thermodynamic properties of extended systems requires performing phonon calculations for large supercells.[19] These are computationally expensive, especially for structures with low symmetry as for example submonolayer coverages (Scheme 1c, d, e) and a more efficient approach is desirable. Such an alternative can be found in the *ab initio* thermodynamics approach as recently reviewed by Reuter.[20,21,22] Here, thermodynamic contributions from the surface are neglected in a first approximation. In order to validate this approach, careful comparison with full phonon calculations and experiment will be carried out. We will focus on the temperature range between T = 300 and 1500 K, typical for growth experiments.[23]

In section 2 we describe the computational methods employed. In section 3 we present the results from *ab initio* thermodynamics (AITD), explicit phonon (EP) and interpolated phonon (IP) computations, leading to a description of coverage dependence on temperature, which we compare to the experimental results described above. We aim at finding the most efficient computational approach to describe the extend of hydrogen coverage of Si(001) under MOVPE conditions.

## II. METHODS

**A Computational details**

Electronic structure calculations based on density functional theory (DFT) were carried out with the Vienna ab-initio simulation package (VASP)[24] version 5.3.5. Structures were optimized using the PBE (Perdew-Becke-Ernzerhof) functional[25] based on the generalized gradient approximation and adding a semiempirical dispersion correction scheme by Grimme et al. (PBE-D3(BJ))[26]. For the D3-scheme, the cutoff radius used is 50.2 Å, and the radius to determine the coordination number was set to 21.2 Å. Additionally, the hybrid functional HSE06-D3(BJ)[27] was used with D3 parameters of the PBE0 functional. The D3 parameters are summarized in Table S1[28] and are taken from Ref. 26b. A plane wave basis set combined with the projector augmented wave method[29] and an energy cutoff of 350 eV was used. The reciprocal space was sampled using a Γ-centered Monkhorst-Pack k-mesh[30] with a (4 2 1)-division for the Si(001) surface for a supercell containing four dimers (Scheme 1).

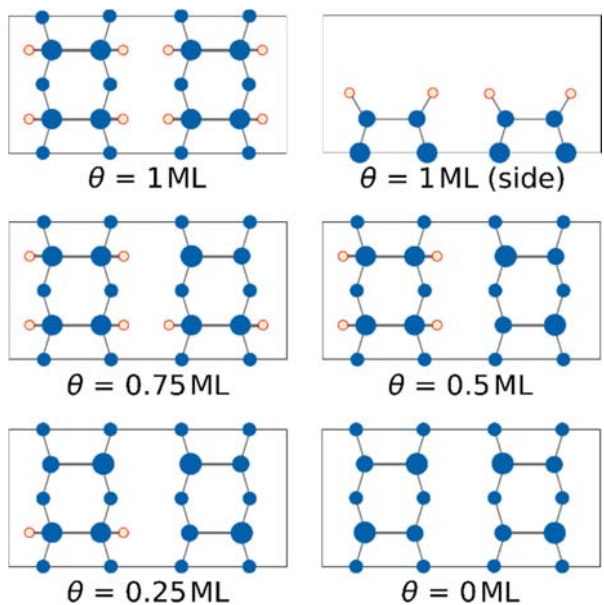

SCHEME 1. Supercells of the Si(001) surface with different coverages $\theta$ used in this study.

Electronic energies were converged to $10^{-5}$ eV while forces for optimizations were converged to $10^{-2}$ eV/Å. For phonon calculations, the Phonopy 1.8.2 code[31,32] was used to generate supercells and compute thermodynamic properties from forces computed with VASP. Throughout, asymmetric slabs with eight layers of silicon atoms were used (six and ten layers for convergence study), where the two bottom layers where kept fixed at bulk positions and saturated with hydogen atoms (frozen double layer approximation). For phonon calculations, (2x2) supercells ((3x3) cells for convergence study) containing 16 dimers were used. A Γ-centered q-mesh with (8 4 1)-division was applied in the determination of the phonon density of states. The theoretically optimized lattice constant of Si was used ($a$ = 5.421 Å; exp.: 5.415 Å)[33]. All thermodynamic correction terms were derived with the computationally more efficient PBE-D3 functional.

The AITD approach used here follows the one described in Ref. 20. While the formulae are derived to give Helmholtz energies, we neglect the volume change of the solid surface and will report Gibbs Energies ($\Delta G$) throughout this manuscript. Thus, the Gibbs Energy change upon dehydrogenation is defined as

$$\Delta G = \frac{N_H}{A} \times \left(\Delta E + \Delta \mu + \frac{1}{2} RT \times \ln\left(\frac{p}{p°}\right)\right). \tag{1}$$

Here, $N_H$ is the number of desorbing hydrogen atoms per unit cell, $A$ is the surface area of the unit cell, $R$ is the universal gas constant, $T$ is the temperature, $p$ is the hydrogen partial pressure and $p°$ is the standard pressure (1013 mbar); the electronic energy difference $\Delta E$ (2) is also called binding energy

$$\Delta E = \frac{E_\theta + \frac{N_H}{2} \times E_{H_2} - E_{1ML}}{N_H}, \tag{2}$$

where $E_\theta$ is the energy of a cell with coverage $\theta$, $E_{H_2}$ the energy of a hydrogen molecule, and $E_{1\,ML}$ the energy of a cell with $\theta$ = 1 ML (see above). Finally, $\Delta \mu$ is the change in chemical potential. In the simplest approximation, it is the chemical potential required to remove one hydrogen atom from a gas phase reservoir. These values are tabulated and the NIST (National Institute of Standards and Technology) offers parameters that can be used with the Shomate equations (3) to obtain accurate enthalpy and entropy values of hydrogen in the gas phase:[34]

$$\frac{H-H°_{RT}}{kJ \times mol^{-1}} = A \times t + B \times \frac{t^2}{2} + C \times \frac{t^3}{3} + D \times \frac{t^4}{4} - \frac{E}{t} + F, \tag{3a}$$

$$\frac{S}{J \times K^{-1} \times mol^{-1}} = A \times \ln(t) + B \times t + C \times \frac{t^2}{2} + D \times \frac{t^3}{3} - \frac{E}{2t^2} + G, \tag{3b}$$

where, $t = \frac{T}{1000K}$. The parameters used in this work are given in Table I.

Another term arises since for any coverage $0 < \theta < 1$ several possibilities for distributing the adsorbate atoms on the adsorption sites in the cell exist. This leads to the configurational entropy term

$$S_{conf} = -nR[(1-\theta)\ln(1-\theta) + \theta \ln \theta], \tag{4}$$

where $n$ is the number of possible adsorption sites.

In the EP approach, the Free Energy difference is computed analogous to the electronic energy difference, but thermodynamic corrections from phonon calculations are added to the electronic energies from DFT runs. These are computed via the supercell approach, where the phonon modes are determined by dislocating atoms and computing the corresponding force constants. The force constants are then further used to determine the phonon density of states (DOS) and thermodynamic corrections.[32] The thermodynamic correction term for hydrogen is taken from the NIST.

TABLE I. Parameters used for the calculation of $\mu(H_2)$ with the Shomate equation (eq. (3)).[34]

|   | 298 – 1000 K | 1000 – 2500 K |
|---|---|---|
| A | 33.0662 | 18.5631 |
| B | -11.3634 | 12.2574 |
| C | 11.4328 | -2.85979 |
| D | -2.77287 | 0.26824 |
| E | -0.15856 | 1.97799 |
| F | -9.9808 | -1.14744 |
| G | 172.708 | 156.288 |

The temperature range for the thermodynamic investigations is 300 – 1500 K, subdivided in 10 K-steps. For the assessment of the approaches outlined above, a typical pressure for MOVPE growth has been used ($p$ = 50 mbar). Furthermore, $p$ was varied up to atmospheric pressure to elucidate the influence of this parameter on the thermodynamic data.

Cartesian coordinates and total energies of all compounds discussed in the text are available in the supplementary material.[28]

## B Convergence studies

For the phonon computations, convergence with respect to slab thickness and supercell size had to be ensured. The convergence of the phonon DOS was used as criterion (shown in Figure S1 in the supplemental material[28]). The phonon DOS for (2x2) and (3x3) supercells are very similar, thus the smaller cell is sufficient. The peaks in the DOS associated with Si-H vibrations (> 85 cm$^{-1}$) are invariant to changes in the slab thickness from six to ten layers, while regions reflecting lattice vibrations (< 85 cm$^{-1}$) of the silicon slab change in intensity according to the increased number of Si atoms for thicker slabs. Vibrations at low frequencies contribute significantly to the entropy corrections. For the relative energies we report here, there will be a significant error compensation regarding this term. Thus, a slab with eight layers of Si atoms and (2x2) supercells for the phonon calculations were used here.

## III. RESULTS AND DISCUSSION

Desorption energies ($\Delta E$) for cleavage of $H_2$ from fully hydrogen-covered H/Si(001) (Scheme 1a, 1b) to sub-monolayer coverages $\theta$ = 0.75, 0.5, 0.25 ML (Scheme 1c-e) and the clean surface ($\theta$ = 0 ML, Scheme 1f) are given in Table 2 derived with the PBE-D3 and HSE06-D3 functionals according to the general desorption reaction (5):

$$[Si - H] \rightarrow [Si_\theta] + \frac{1}{2} H_2. \tag{5}$$

We use the sign convention of positive $\Delta E$ values indicating an energetically unfavorable reaction. It is found that the reaction energies for the desorption reaction (5) are positive for all coverages $\theta$ (Table II). The desorption energy depends slightly on the final coverage, with desorption from $\theta$ = 1 ML to $\theta$ = 0.75 ML being the most endo-energetic process. However, this is probably a cell-size effect since the short distance between the translational images can lead to spurious interactions due to the periodic boundary conditions employed. To eliminate this unphysical effect, an average desorption energy is given in Table 2 which will be used for the rest of the study. Desorption energies computed with HSE06-D3 are roughly 15 kJ mol$^{-1}$ higher compared to PBE-D3 results. The values computed with HSE06-D3 agree slightly better with a range of other reported chemisorption energies,[11] thus this functional is used further-on for $\Delta E$.

TABLE II. Desorption energies per H atom ($\Delta E$, see eq. (5)) from a monolayer to various coverages computed with PBE-D3 and HSE06-D3.

| $\theta_{fin}$ | 0.75 | 0.50 | 0.25 | 0.00 | $\Delta E_{av}$[a] |
|---|---|---|---|---|---|
| PBE-D3 | 92.9 | 88.8 | 90.1 | 88.7 | 90.1 |
| HSE06-D3 | 108.2 | 104.2 | 105.4 | 104.0 | 105.5 |

[a]averaged value for all coverages

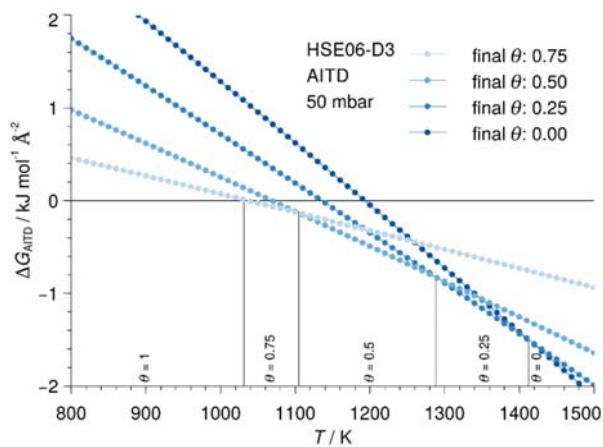

FIGURE 1. *Ab initio* thermodynamics (AITD) approach: $\Delta G_{AITD}$ values according to reaction (5) based on HSE06-D3 energies at $p = 50$ mbar.

Thermodynamic correction of the electronic desorption energies was first carried out with the AITD approach leading to Gibbs energies of desorption $\Delta G_{AITD}$ presented in Figure 1. The fully covered surface is stable up to ca. 1025 K, $\theta = 0.75$ ML is stable for a temperature range of about 1025 - 1100 K, followed by a range of $T = 1100$-1280 K with $\theta = 0.5$ ML. From 1280 – 1400 K, $\theta = 0.25$ ML is the most stable range and the pristine surface is only found above ca. 1410 K. The broad temperature range for $\theta = 0.5$ ML can be attributed to the configurational entropy term, which is largest for this coverage. Neglect of this term would lead to prediction of a complete desorption of the adsorbed hydrogen atoms at $T = 1040$ K without intermediate coverages, which is unphysical. Since the temperatures studied are rather high, the neglect of surface phonons is a severe approximation and needs to be checked in the next step by taking into account thermodynamic correction terms for the surface.

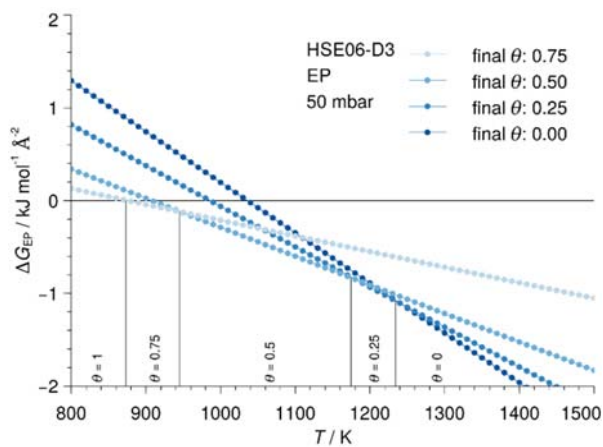

FIGURE 2. Explicit phonon (EP) approach: $\Delta G_{EP}$ values according to reaction (5) based on HSE06-D3 energies at $p = 50$ mbar.

Deriving the phonon spectrum for the surface (EP approach) leads to the Gibbs energies of desorption $\Delta G_{EP}$ presented in Figure 2. While the qualitative picture remains the same compared to the AITD approach (Figure 1), the temperature for on-set of hydrogen desorption is reduced by approximately 200 K and $\theta = 0.75$ ML is found to be more stable than $\theta = 1$ ML at $T = 875$ K. This difference between EP and AITD approaches points toward a non-negligible thermodynamic stabilization of the partially hydrogen-covered surface (Equation (5)). Notably, the temperature range where $\theta = 0.25$ ML is predicted to be stable is significantly narrower compared to the AITD approach (ca. 60 K vs. 130 K for AITD). The pristine surface is reached at ca. 1240 K based on the EP approach.

Since deriving phonon spectra is computationally very demanding, especially for intermediate coverage structures of low symmetry, a more feasible alternative is desirable. The thermodynamic correction terms of intermediate coverage structures can be linearly interpolated between the initial and final structures of full and no coverage:

$$\mu(\theta) = \theta * \mu_{\theta=1} + (1-\theta) * \mu_{\theta=0}. \tag{6}$$

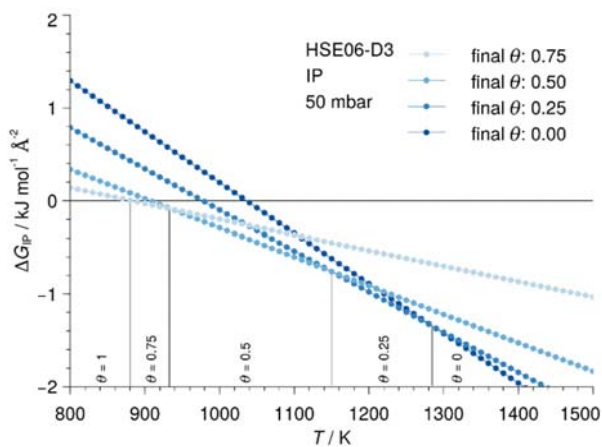

FIGURE 3. Interpolated phonon (IP) approach: $\Delta G_{IP}$ values according to reaction (5) based on HSE06-D3 energies at $p$ = 50 mbar.

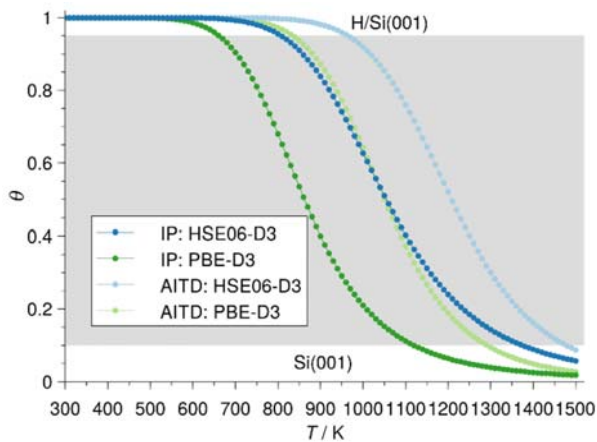

FIGURE 4. Temperature dependence of coverage θ for IP and AITD approach. Binding energies computed with PBE-D3 and HSE06-D3. The grey-shaded area indicates the range of the graph with partially hydrogenated surface (0.95 > $\theta$ > 0.1 ML).

Since there are eight possible adsorption positions for hydrogen atoms in the supercell, the amount of hydrogen atoms transferred can be expressed as

$$N_H = (1 - \theta_{fin}) \times 8. \tag{7}$$

The $\Delta G_{IP}$ values obtained with this IP approach are presented in Figure 3. The agreement between the sophisticated EP approach and the much simpler IP ansatz is surprisingly good. Small differences in the explicitly computed and interpolated chemical potential lead to different slopes of the Gibbs energy and thus to slightly different temperatures of coverage change. But the transition temperatures between the different coverages differ by only up to 40 K. However, these differences can be neglected without changing conclusions.

For a real surface any intermediate coverage is possible, thus it is convenient to be able to treat coverage as a continuous variable. Using the IP approach and the averaged electronic energy difference, $\Delta G_{IP}$ can be expressed in a form that leaves $\theta$ as the only remaining variable:

$$\Delta G_{IP} = \frac{(1-\theta)8}{A}\left(\Delta E + \Delta\mu(\theta, T) + \frac{1}{2}RT\ln\left(\frac{p}{p°}\right)\right) + \frac{8}{A}RT[\theta\ln\theta + (1-\theta)\ln(1-\theta)]. \tag{8}$$

For each temperature T, the coverage which minimizes $\Delta G_{IP}$ is the one which should be observed experimentally. This leads to a particularly instructive representation shown in Figure 4. The Figure here includes results from both functionals used in this study: PBE-D3 and HSE06-D3. This (quasi) continuous formulation of temperature dependent coverage predicts a small initial desorption which becomes larger for higher temperatures in line with the known desorption dynamics of the system, e. g. visible in TPD experiments.[15] The change of desorption with respect to temperature is highest for coverages in the vicinity of $\theta$ = 0.5 ML. For high temperatures ($T$ > 1400 K), the pristine surface is reached. A small residual coverage is predicted even for high temperatures. It should be noted that the coverage trend for the IP approach based on HSE06-D3 and the AITD approach based on PBE-D3 coincide to a large extent. Thus, there is obviously error compensation in the description of the electronic and thermodynamic effects leading to the apparent conclusion that the simpler functional (PBE-D3) with the simpler thermodynamic approximation (AITD) gives accurate results.

In order to check if refining the AITD approach by applying the Einstein model[35] to the vibrational modes of adsorbed hydrogen is viable, the Free Energy change was determined for complete desorption. The phonon DOS of the hydrogenated and pristine surface is shown in Figure 5(a) (the phonon dispersion relation is given in Figure S2[28]). The most prominent changes occur at 95 and 335 cm$^{-1}$, where the phonon DOS is determined by Si-H vibrations. Some smaller changes can be seen at 75 and 45 cm$^{-1}$. Only the two larger contributions are taken into account for this model. For a monoatomic solid, the Gibbs Energy within the Einstein model is given by[36]

$$G(\omega, T) = -aRT \, \log\left(\frac{\exp\left(\frac{\hbar\omega}{2kT}\right)}{1-\exp\left(\frac{\hbar\omega}{kT}\right)}\right). \tag{9}$$

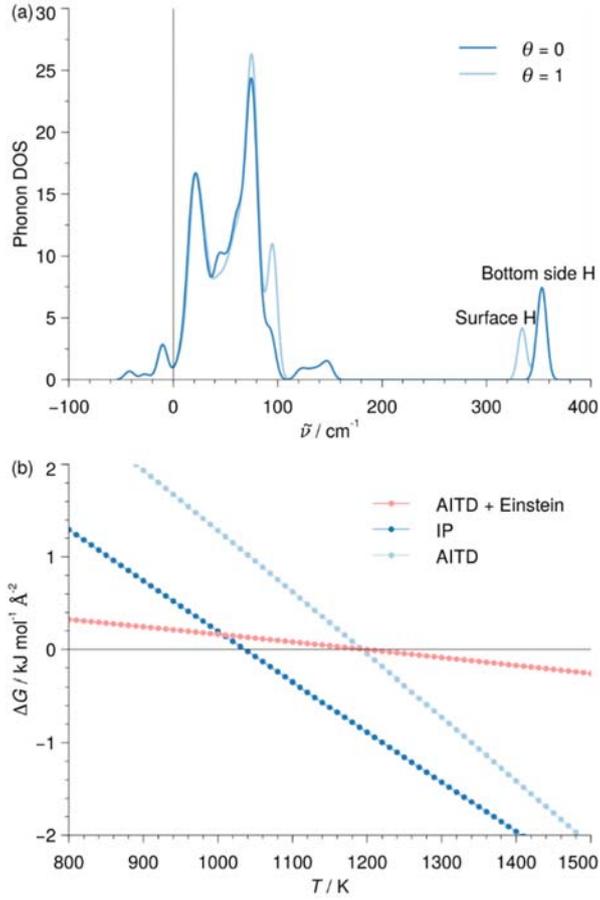

FIGURE 5. (a) Phonon density of states (DOS) of hydrogenated and pristine Si surface. The negative frequencies stem from the frozen bottom layers. (b) Free energy of complete desorption for Einstein model, IP and AITD approach in comparison. The electronic energies were computed with HSE06-D3.

A monoatomic solid has three degrees of freedom (i. e. phonon branches), thus a = 3 in equation (9). The contribution of a single phonon branch for systems with more than one frequency can be estimated by choosing a = 1 equation (9). For every adsorbed H atom, one phonon branch at 335 cm$^{-1}$ and two branches at 95 cm$^{-1}$ show up. Thus, the Free Energy correction for the hydrogenated surface is

$$G_{\theta=1}(T) = N\left(G(335\ \text{cm}^{-1}, T) + 2\ G(95\ \text{cm}^{-1}, T)\right), \tag{10}$$

with $N$ being the number of adsorbed H atoms per unit cell, here $N = 8$. The resulting Free Energy curve is shown in Figure 5(b). While the temperature, at which desorption is predicted, is similar to the AITD approach, it deviates from the more accurate IP approach and the curve is much more flat. We conclude that the Einstein model is not accurate enough to describe the temperature dependence of hydrogen coverage for the system at hand.

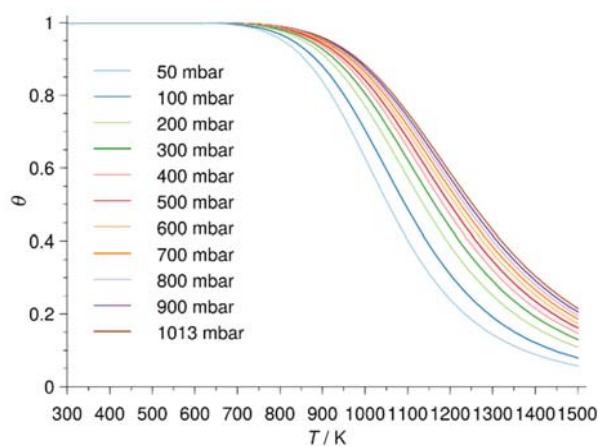

FIGURE 6. Pressure and temperature dependence of coverage for interpolated phonon approach based on HSE06-D3 binding energies.

Finally, we investigated the pressure dependence of surface coverage $p$ = 50 – 1013 mbar as shown in Figure 6. Higher pressure shifts the coverage for a given temperature to higher values as expected from Le Châtelier's principle. In agreement with the logarithmic dependence of Gibbs energies on pressure,[37] the effect of pressure change decreases with increasing total pressure. The overall effect of pressure change is rather small and can thus be neglected in studies of MOVPE reactions where the pressure changes are on the order of several tenth of mbar.

In comparison with the experimental results reported in Ref. [17], the data produced by the most accurate computations (HSE06-D3 energies and EP corrections) give the best agreement of the approaches tested here and a very good agreement in absolute terms. The experimentally observed transition from pristine to hydrogen-covered surface under MOVPE conditions occurs around $T = 1070$ K.[17] At this temperature, the IP method predicts a significant dehydrogenation degree ($\theta = 0.5$ ML). Thus, the temperature where the pristine surface becomes predominant over the hydrogenated phase is predicted consistently. The on-set of hydrogen desorption from the fully covered surface occurs earlier in the theoretical description ($T \approx 700$ K, Figure 4) compared to the high-pressure experiments. But it is in very good agreement with previous TPD studies (700-780 K).[15,16] In addition, the temperature range where desorption occurs seems to be narrower in experiment compared to our results. For atmospheric pressure, the coverage at low temperatures is underestimated, while it is overestimated at high temperatures, as is the case for MOVPE pressure.

## IV.  CONCLUSIONS

The desorption of hydrogen from the fully hydrogenated H/Si(001) surface under thermodynamic equilibrium conditions has been studied with a combination of density functional theory, *ab initio* thermodynamics and phonon computations. In comparison to available experimental data, the most accurate approach tested here (hybrid functional with dispersion correction for the electronic energy and calculation of the phonon spectrum of the system) delivers good agreement regarding the on-set temperature for desorption. Additionally, the temperature range where desorption was observed experimentally is reproduced. However, the desorption curve predicted here is less steep compared to experiment. It might be worthwhile to study the persistence of residual hydrogen at high temperatures and the initial desorption at low temperatures by experimental means. In the relevant temperature range for MOVPE, our computations predict a significant ($\theta < 0.95$ ML) desorption starting at approximately $T = 800$ K. A nearly pristine surface ($\theta < 0.1$ ML) is reached at approximately $T = 1400$ K. Beside the demanding computation of explicit phonons, the much more efficient interpolated phonon can be used in the future as well. *Ab initio* thermodynamics with Einstein corrections are less accurate for the system at hand but might be suitable for systems with heavier atoms.

This study now enables us to investigate adsorption and surface reactivity of organic and inorganic precursor molecules relevant for thin-film growth with metal organic vapor phase epitaxy with an accurate representation of the surface at the simulation temperature.

**ACKNOWLEDGEMENTS**

This work was supported by the Deutsche Forschungsgemeinschaft via Research Training Group 1782 "Functionalization of Semiconductors". We thank Prof. Karsten Reuter (Munich) for his hospitality during a research visit by P.R. and discussions. We thank Prof. Axel Groß (Ulm), Prof. Wolfgang Stolz (Marburg) and Prof. Michael Dürr (Gießen) for discussions. Computational resources were provided by the Hochschulrechenzentrum Marburg, the LOEWE-CSC Frankfurt and the HLRS Stuttgart.

# Supporting Information for

# Extend of hydrogen coverage of Si(001) under chemical vapor deposition conditions from *ab initio* approaches

Phil Rosenow, Ralf Tonner

**Table S1.** D3-Parameters used for PBE-D3 and HSE06-D3 calculations (HSE-parameters are those for PBE0). The expression for the dispersion correction is given according to S. Grimme, S. Ehrlich, and L. Goerigk, J. Comp. Chem. 32, 1456 (2011) by :

$$E_{\text{disp}}^{\text{D3}} = -\frac{1}{2} \sum_{A \neq B} s_6 \frac{C_6^{AB}}{R_{AB}^6 + [f(R_{AB}^0)]^6} + s_8 \frac{C_8^{AB}}{R_{AB}^8 + [f(R_{AB}^0)]^8}$$

$$\text{with } f(R_{AB}^0) = a_1 R_{AB}^0 + a_2 \text{ and } R_{AB}^0 = \sqrt{\frac{C_8^{AB}}{C_6^{AB}}}.$$

The parameter $s_6$ is set to unity.

|       | s8     | a1     | a2     |
|-------|--------|--------|--------|
| PBE   | 0.7875 | 0.4289 | 4.4407 |
| HSE06 | 1.2177 | 0.4145 | 4.8593 |

**Table S2.** Total energies of used unit cells in eV.

| System | PBE-D3 | HSE06-D3 |
|---|---|---|
| molecular $H_2$ | -6.74816899 | -7.80958547 |
| Si(001) | -405.38169351 | -469.43016295 |
| $\theta$ = 0.25 ML | -413.88348356 | -479.30568130 |
| $\theta$ = 0.50 ML | -422.55333648 | -489.35290438 |
| $\theta$ = 0.75 ML | -431.05702666 | -499.23812543 |
| $\theta$ = 1 ML (H/Si(001)) | -439.73108245 | -509.29140990 |

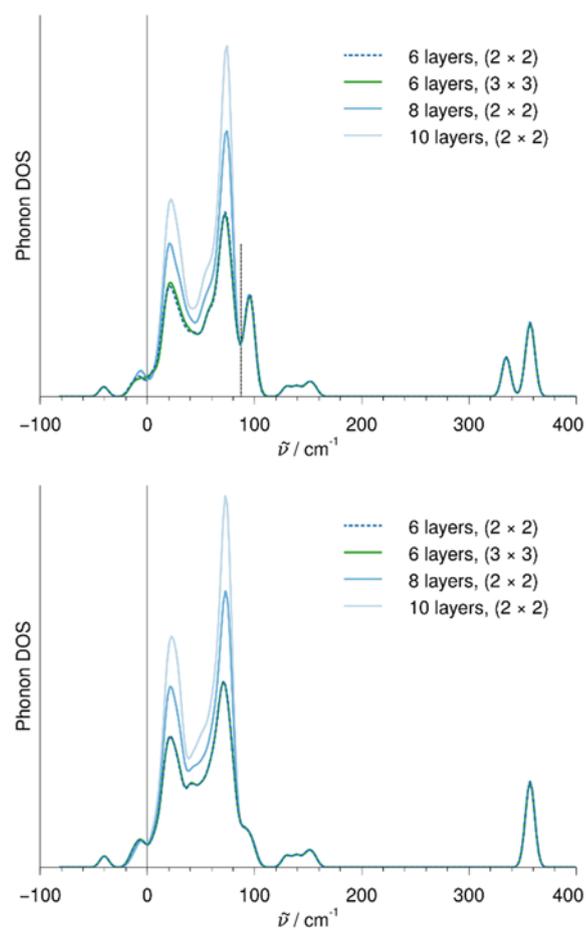

**Figure S1**. Phonon densities of states (DOS) of hydrogenated (top) and pristine (bottom) surface cells with varying number of layers and supercell size for 6 layer cell. The vertical dashed line in the top figure shows the on-set of Si-H vibrations (to higher vibrational energies).

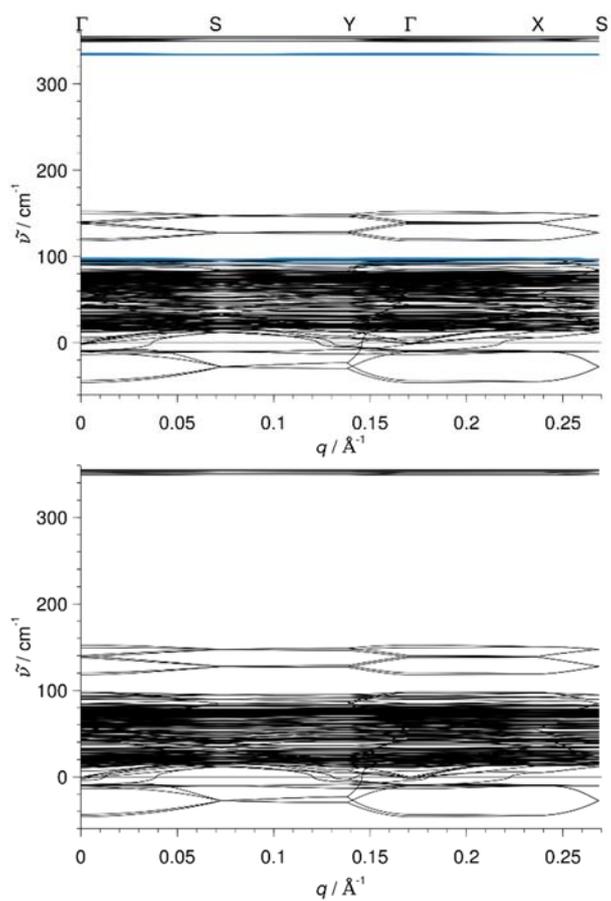

**Figure S2**. Phonon dispersion relation for the monohydrogenated (top) and pristine (bottom) surface. Branches from surface Si-H vibrations are colored in blue.

**Optimized coordinates of structures investigated in fractional coordinates (VASP format)**

```
Pristine surface
  1.00000000000000
    7.6664519310000001    0.0000000000000000    0.0000000000000000
    0.0000000000000000   15.3329038620000002    0.0000000000000000
    0.0000000000000000    0.0000000000000000   24.3945007324000009
  Si   H
  64   16
Selective dynamics
Direct
 0.0000000000000000  0.1877089947222349  0.4654003536532387   T  T  T
 0.0000000000000000  0.6658508867043397  0.4958955467171131   T  T  T
 0.5000000000000000  0.1658508867043390  0.4958955467171131   T  T  T
 0.5000000000000000  0.6877089947222412  0.4654003536532387   T  T  T
 0.0000000000000000  0.3341490982956495  0.4958955467171131   T  T  T
 0.0000000000000000  0.8122909902777469  0.4654003536532387   T  T  T
 0.5000000000000000  0.3122909902777528  0.4654003536532387   T  T  T
 0.5000000000000000  0.8341490982956484  0.4958955467171131   T  T  T
 0.2664989330698293  0.1316266534045507  0.4348590053621761   T  T  T
 0.2335010669301723  0.6316266534045524  0.4348590053621832   T  T  T
 0.7335010669301731  0.1316266534045507  0.4348590053621761   T  T  T
 0.7664989330698269  0.6316266534045524  0.4348590053621761   T  T  T
 0.2335010669301723  0.3683733465954504  0.4348590053621761   T  T  T
 0.2664989330698293  0.8683733465954476  0.4348590053621761   T  T  T
 0.7664989330698269  0.3683733465954504  0.4348590053621832   T  T  T
 0.7335010669301731  0.8683733465954476  0.4348590053621832   T  T  T
 0.2522982880860112  0.0000000000000000  0.3837999529132446   T  T  T
 0.2477017119139890  0.5000000000000000  0.3837999529132446   T  T  T
 0.7477017119139867  0.0000000000000000  0.3837999529132446   T  T  T
 0.7522982880860133  0.5000000000000000  0.3837999529132446   T  T  T
 0.2500000000000000  0.2500000000000000  0.3718240620545487   T  T  T
 0.2500000000000000  0.7500000000000000  0.3718240620545487   T  T  T
 0.7500000000000000  0.2500000000000000  0.3718240620545487   T  T  T
 0.7500000000000000  0.7500000000000000  0.3718240620545487   T  T  T
 0.0000000000000000  0.0000000000000000  0.3273507331796979   T  T  T
 0.0000000000000000  0.5000000000000000  0.3255642376310458   T  T  T
 0.5000000000000000  0.0000000000000000  0.3255642376310458   T  T  T
 0.5000000000000000  0.5000000000000000  0.3273507331796979   T  T  T
 0.0000000000000000  0.2511949888625770  0.3175799769998215   T  T  T
 0.0000000000000000  0.7488050111374266  0.3175799769998215   T  T  T
 0.5000000000000000  0.2488050111374236  0.3175799769998215   T  T  T
 0.5000000000000000  0.7511949888625734  0.3175799769998215   T  T  T
 0.0000000000000000  0.1219759002904503  0.2672311982592552   T  T  T
 0.0000000000000000  0.6215656210789557  0.2660067948723296   T  T  T
 0.5000000000000000  0.1215656210789568  0.2660067948723296   T  T  T
```

```
  0.5000000000000000  0.6219759002904456  0.2672311982592552  T  T  T
  0.0000000000000000  0.3784343789210411  0.2660067948723296  T  T  T
  0.0000000000000000  0.8780240997095544  0.2672311982592552  T  T  T
  0.5000000000000000  0.3780240997095571  0.2672311982592552  T  T  T
  0.5000000000000000  0.8784343789210443  0.2660067948723296  T  T  T
  0.2491871700502964  0.1237687871929941  0.2111414992167098  T  T  T
  0.2508128299496942  0.6237687871929796  0.2111414992167098  T  T  T
  0.7508128299497018  0.1237687871929941  0.2111414992167098  T  T  T
  0.7491871700502982  0.6237687871929796  0.2111414992167098  T  T  T
  0.2508128299496942  0.3762312128069903  0.2111414992167098  T  T  T
  0.2491871700502964  0.8762312128070062  0.2111414992167098  T  T  T
  0.7491871700502982  0.3762312128069903  0.2111414992167098  T  T  T
  0.7508128299497018  0.8762312128070062  0.2111414992167098  T  T  T
  0.2500000000000000  0.0000000000000000  0.1555555609999999  F  F  F
  0.2500000000000000  0.5000000000000000  0.1555555609999999  F  F  F
  0.7500000000000000  0.0000000000000000  0.1555555609999999  F  F  F
  0.7500000000000000  0.5000000000000000  0.1555555609999999  F  F  F
  0.2500000000000000  0.2500000000000000  0.1555555609999999  F  F  F
  0.2500000000000000  0.7500000000000000  0.1555555609999999  F  F  F
  0.7500000000000000  0.2500000000000000  0.1555555609999999  F  F  F
  0.7500000000000000  0.7500000000000000  0.1555555609999999  F  F  F
  0.0000000000000000  0.0000000000000000  0.1000000009999980  F  F  F
  0.0000000000000000  0.5000000000000000  0.1000000009999980  F  F  F
  0.5000000000000000  0.0000000000000000  0.1000000009999980  F  F  F
  0.5000000000000000  0.5000000000000000  0.1000000009999980  F  F  F
  0.0000000000000000  0.2500000000000000  0.1000000009999980  F  F  F
  0.0000000000000000  0.7500000000000000  0.1000000009999980  F  F  F
  0.5000000000000000  0.2500000000000000  0.1000000009999980  F  F  F
  0.5000000000000000  0.7500000000000000  0.1000000009999980  F  F  F
  0.0000000000000000  0.0788012370000004  0.0649772224900005  F  F  F
  0.0000000000000000  0.5788012370000004  0.0649772224900005  F  F  F
  0.5000000000000000  0.0788012370000004  0.0649772224900005  F  F  F
  0.5000000000000000  0.5788012370000004  0.0649772224900005  F  F  F
  0.0000000000000000  0.1711987679999964  0.0649772224900005  F  F  F
  0.0000000000000000  0.6711987679999964  0.0649772224900005  F  F  F
  0.5000000000000000  0.1711987679999964  0.0649772224900005  F  F  F
  0.5000000000000000  0.6711987679999964  0.0649772224900005  F  F  F
  0.0000000000000000  0.3288012370000004  0.0649772224900005  F  F  F
  0.0000000000000000  0.8288012370000004  0.0649772224900005  F  F  F
  0.5000000000000000  0.3288012370000004  0.0649772224900005  F  F  F
  0.5000000000000000  0.8288012370000004  0.0649772224900005  F  F  F
  0.0000000000000000  0.4211987629999996  0.0649772224900005  F  F  F
  0.0000000000000000  0.9211987629999996  0.0649772224900005  F  F  F
  0.5000000000000000  0.4211987629999996  0.0649772224900005  F  F  F
  0.5000000000000000  0.9211987629999996  0.0649772224900005  F  F  F
```

  Hydrogenated surface
   1.00000000000000
     7.6664519310000001    0.0000000000000000    0.0000000000000000

```
     0.0000000000000000    15.3329038620000002    0.0000000000000000
     0.0000000000000000     0.0000000000000000   24.3945007324000009
  Si   H
   64   24
Selective dynamics
Direct
  0.0000000000000000  0.1715648153692868  0.4838406285918566   T   T   T
  0.0000000000000000  0.6715648153692868  0.4838406285918566   T   T   T
  0.5000000000000000  0.1715648153692868  0.4838406285918566   T   T   T
  0.5000000000000000  0.6715648153692868  0.4838406285918566   T   T   T
  0.0000000000000000  0.3284351696307013  0.4838406285918566   T   T   T
  0.0000000000000000  0.8284351696307013  0.4838406285918566   T   T   T
  0.5000000000000000  0.3284351696307013  0.4838406285918566   T   T   T
  0.5000000000000000  0.8284351696307013  0.4838406285918566   T   T   T
  0.2500000000000000  0.1319112315587390  0.4337767662223371   T   T   T
  0.2500000000000000  0.6319112315587390  0.4337767662223371   T   T   T
  0.7500000000000000  0.1319112315587390  0.4337767662223371   T   T   T
  0.7500000000000000  0.6319112315587390  0.4337767662223371   T   T   T
  0.2500000000000000  0.3680887684412610  0.4337767662223371   T   T   T
  0.2500000000000000  0.8680887684412610  0.4337767662223371   T   T   T
  0.7500000000000000  0.3680887684412610  0.4337767662223371   T   T   T
  0.7500000000000000  0.8680887684412610  0.4337767662223371   T   T   T
  0.2500000000000000  0.0000000000000000  0.3830810445681507   T   T   T
  0.2500000000000000  0.5000000000000000  0.3830810445681507   T   T   T
  0.7500000000000000  0.0000000000000000  0.3830810445681507   T   T   T
  0.7500000000000000  0.5000000000000000  0.3830810445681507   T   T   T
  0.2500000000000000  0.2500000000000000  0.3725801948939065   T   T   T
  0.2500000000000000  0.7500000000000000  0.3725801948939065   T   T   T
  0.7500000000000000  0.2500000000000000  0.3725801948939065   T   T   T
  0.7500000000000000  0.7500000000000000  0.3725801948939065   T   T   T
  0.0000000000000000  0.0000000000000000  0.3258920101530194   T   T   T
  0.0000000000000000  0.5000000000000000  0.3258920101530194   T   T   T
  0.5000000000000000  0.0000000000000000  0.3258920101530194   T   T   T
  0.5000000000000000  0.5000000000000000  0.3258920101530194   T   T   T
  0.0000000000000000  0.2500000000000000  0.3186114950424468   T   T   T
  0.0000000000000000  0.7500000000000000  0.3186114950424468   T   T   T
  0.5000000000000000  0.2500000000000000  0.3186114950424468   T   T   T
  0.5000000000000000  0.7500000000000000  0.3186114950424468   T   T   T
  0.0000000000000000  0.1224082624176432  0.2667630817947924   T   T   T
  0.0000000000000000  0.6224082624176432  0.2667630817947924   T   T   T
  0.5000000000000000  0.1224082624176432  0.2667630817947924   T   T   T
  0.5000000000000000  0.6224082624176432  0.2667630817947924   T   T   T
  0.0000000000000000  0.3775917375823568  0.2667630817947924   T   T   T
  0.0000000000000000  0.8775917375823568  0.2667630817947924   T   T   T
  0.5000000000000000  0.3775917375823568  0.2667630817947924   T   T   T
  0.5000000000000000  0.8775917375823568  0.2667630817947924   T   T   T
  0.2500000000000000  0.1238842470464974  0.2112190832977277   T   T   T
  0.2500000000000000  0.6238842470464974  0.2112190832977277   T   T   T
  0.7500000000000000  0.1238842470464974  0.2112190832977277   T   T   T
  0.7500000000000000  0.6238842470464974  0.2112190832977277   T   T   T
```

```
0.2500000000000000  0.3761157529534884  0.2112190832977277  T  T  T
0.2500000000000000  0.8761157529534884  0.2112190832977277  T  T  T
0.7500000000000000  0.3761157529534884  0.2112190832977277  T  T  T
0.7500000000000000  0.8761157529534884  0.2112190832977277  T  T  T
0.2500000000000000  0.0000000000000000  0.1555555609999999  F  F  F
0.2500000000000000  0.5000000000000000  0.1555555609999999  F  F  F
0.7500000000000000  0.0000000000000000  0.1555555609999999  F  F  F
0.7500000000000000  0.5000000000000000  0.1555555609999999  F  F  F
0.2500000000000000  0.2500000000000000  0.1555555609999999  F  F  F
0.2500000000000000  0.7500000000000000  0.1555555609999999  F  F  F
0.7500000000000000  0.2500000000000000  0.1555555609999999  F  F  F
0.7500000000000000  0.7500000000000000  0.1555555609999999  F  F  F
0.0000000000000000  0.0000000000000000  0.1000000009999980  F  F  F
0.0000000000000000  0.5000000000000000  0.1000000009999980  F  F  F
0.5000000000000000  0.0000000000000000  0.1000000009999980  F  F  F
0.5000000000000000  0.5000000000000000  0.1000000009999980  F  F  F
0.0000000000000000  0.2500000000000000  0.1000000009999980  F  F  F
0.0000000000000000  0.7500000000000000  0.1000000009999980  F  F  F
0.5000000000000000  0.2500000000000000  0.1000000009999980  F  F  F
0.5000000000000000  0.7500000000000000  0.1000000009999980  F  F  F
0.0000000000000000  0.0788012370000004  0.0649772224900005  F  F  F
0.0000000000000000  0.5788012370000004  0.0649772224900005  F  F  F
0.5000000000000000  0.0788012370000004  0.0649772224900005  F  F  F
0.5000000000000000  0.5788012370000004  0.0649772224900005  F  F  F
0.0000000000000000  0.1711987679999964  0.0649772224900005  F  F  F
0.0000000000000000  0.6711987679999964  0.0649772224900005  F  F  F
0.5000000000000000  0.1711987679999964  0.0649772224900005  F  F  F
0.5000000000000000  0.6711987679999964  0.0649772224900005  F  F  F
0.0000000000000000  0.3288012370000004  0.0649772224900005  F  F  F
0.0000000000000000  0.8288012370000004  0.0649772224900005  F  F  F
0.5000000000000000  0.3288012370000004  0.0649772224900005  F  F  F
0.5000000000000000  0.8288012370000004  0.0649772224900005  F  F  F
0.0000000000000000  0.4211987629999996  0.0649772224900005  F  F  F
0.0000000000000000  0.9211987629999996  0.0649772224900005  F  F  F
0.5000000000000000  0.4211987629999996  0.0649772224900005  F  F  F
0.5000000000000000  0.9211987629999996  0.0649772224900005  F  F  F
0.0000000000000000  0.1367625824061633  0.5414012781483777  T  T  T
0.0000000000000000  0.6367625824061633  0.5414012781483777  T  T  T
0.5000000000000000  0.1367625824061633  0.5414012781483777  T  T  T
0.5000000000000000  0.6367625824061633  0.5414012781483777  T  T  T
0.0000000000000000  0.3632374325938414  0.5414012781483777  T  T  T
0.0000000000000000  0.8632374325938414  0.5414012781483777  T  T  T
0.5000000000000000  0.3632374325938414  0.5414012781483777  T  T  T
0.5000000000000000  0.8632374325938414  0.5414012781483777  T  T  T
```

Coverage 0.25 ML
  1.00000000000000
    7.6664519310000001    0.0000000000000000    0.0000000000000000
    0.0000000000000000   15.3329038620000002    0.0000000000000000

```
     0.0000000000000000    0.0000000000000000   24.3945007324000009
   Si   H
    64    18
Selective dynamics
Direct
  0.0000000000000000  0.1653838455652470  0.4958897101275387   T   T   T
  0.0000000000000000  0.6940889323922135  0.4662630670318215   T   T   T
  0.5000000000000000  0.1875088396217694  0.4656708320214106   T   T   T
  0.5000000000000000  0.6764906200558599  0.4851853922988072   T   T   T
  0.0000000000000000  0.3118099687780388  0.4656872477097657   T   T   T
  0.0000000000000000  0.8391119472420099  0.4936660295183461   T   T   T
  0.5000000000000000  0.3341556037407162  0.4963216535471419   T   T   T
  0.5000000000000000  0.8331123296637347  0.4837830697168647   T   T   T
  0.2339456926514372  0.1311192613493573  0.4349825410061945   T   T   T
  0.2572323257766810  0.6314432445691497  0.4336546724104520   T   T   T
  0.7660543073485595  0.1311192613493573  0.4349825410061945   T   T   T
  0.7427676742233220  0.6314432445691497  0.4336546724104520   T   T   T
  0.2664497631198944  0.3679429105984757  0.4351643668744991   T   T   T
  0.2431291168439805  0.8673988381463660  0.4353491925400985   T   T   T
  0.7335502368801093  0.3679429105984757  0.4351643668744991   T   T   T
  0.7568708831560255  0.8673988381463660  0.4353491925400985   T   T   T
  0.2480296280964208 -0.0008318509271488  0.3841832499901931   T   T   T
  0.2513642921347442  0.4989075522835142  0.3834417660110253   T   T   T
  0.7519703719035787 -0.0008318509271488  0.3841832499901931   T   T   T
  0.7486357078652546  0.4989075522835142  0.3834417660110253   T   T   T
  0.2501822493170657  0.2497289449364082  0.3720946837750896   T   T   T
  0.2525716769932316  0.7520744569743842  0.3722407104363759   T   T   T
  0.7498177506829340  0.2497289449364082  0.3720946837750896   T   T   T
  0.7474283230067535  0.7520744569743842  0.3722407104363759   T   T   T
  0.0000000000000000 -0.0008738363388621  0.3260493560098424   T   T   T
  0.0000000000000000  0.4991521473653823  0.3266510773058592   T   T   T
  0.5000000000000000  0.0001295882951740  0.3274631828839770   T   T   T
  0.5000000000000000  0.5003651222637140  0.3255458505822840   T   T   T
  0.0000000000000000  0.2481240645807009  0.3178678209792197   T   T   T
  0.0000000000000000  0.7498386217162152  0.3192962064078959   T   T   T
  0.5000000000000000  0.2510745067276698  0.3177521389495240   T   T   T
  0.5000000000000000  0.7491309297134374  0.3172091203059392   T   T   T
  0.0000000000000000  0.1207861822575630  0.2663218962108443   T   T   T
  0.0000000000000000  0.6220603812632022  0.2674339355452635   T   T   T
  0.5000000000000000  0.1219594734444661  0.2672887055928457   T   T   T
  0.5000000000000000  0.6217829115227855  0.2658982248070513   T   T   T
  0.0000000000000000  0.3769305202013887  0.2670436686414399   T   T   T
  0.0000000000000000  0.8769824632958707  0.2668203153680050   T   T   T
  0.5000000000000000  0.3783218845570395  0.2662003792911106   T   T   T
  0.5000000000000000  0.8783253050566371  0.2670286825941383   T   T   T
  0.2506117350806036  0.1237497761753201  0.2112631510604167   T   T   T
  0.2489826148160026  0.6237584817213669  0.2111716137691282   T   T   T
  0.7493882649193966  0.1237497761753201  0.2112631510604167   T   T   T
  0.7510173851840045  0.6237584817213669  0.2111716137691282   T   T   T
  0.2494712797759603  0.3761760534273773  0.2111312674542582   T   T   T
```

```
  0.2501631538105983  0.8761362473806615  0.2113161923788887   T   T   T
  0.7505287202240456  0.3761760534273773  0.2111312674542582   T   T   T
  0.7498368461894085  0.8761362473806615  0.2113161923788887   T   T   T
  0.2500000000000000  0.0000000000000000  0.1555555609999999   F   F   F
  0.2500000000000000  0.5000000000000000  0.1555555609999999   F   F   F
  0.7500000000000000  0.0000000000000000  0.1555555609999999   F   F   F
  0.7500000000000000  0.5000000000000000  0.1555555609999999   F   F   F
  0.2500000000000000  0.2500000000000000  0.1555555609999999   F   F   F
  0.2500000000000000  0.7500000000000000  0.1555555609999999   F   F   F
  0.7500000000000000  0.2500000000000000  0.1555555609999999   F   F   F
  0.7500000000000000  0.7500000000000000  0.1555555609999999   F   F   F
  0.0000000000000000  0.0000000000000000  0.1000000009999980   F   F   F
  0.0000000000000000  0.5000000000000000  0.1000000009999980   F   F   F
  0.5000000000000000  0.0000000000000000  0.1000000009999980   F   F   F
  0.5000000000000000  0.5000000000000000  0.1000000009999980   F   F   F
  0.0000000000000000  0.2500000000000000  0.1000000009999980   F   F   F
  0.0000000000000000  0.7500000000000000  0.1000000009999980   F   F   F
  0.5000000000000000  0.2500000000000000  0.1000000009999980   F   F   F
  0.5000000000000000  0.7500000000000000  0.1000000009999980   F   F   F
  0.0000000000000000  0.0788012370000004  0.0649772224900005   F   F   F
  0.0000000000000000  0.5788012370000004  0.0649772224900005   F   F   F
  0.5000000000000000  0.0788012370000004  0.0649772224900005   F   F   F
  0.5000000000000000  0.5788012370000004  0.0649772224900005   F   F   F
  0.0000000000000000  0.1711987679999964  0.0649772224900005   F   F   F
  0.0000000000000000  0.6711987679999964  0.0649772224900005   F   F   F
  0.5000000000000000  0.1711987679999964  0.0649772224900005   F   F   F
  0.5000000000000000  0.6711987679999964  0.0649772224900005   F   F   F
  0.0000000000000000  0.3288012370000004  0.0649772224900005   F   F   F
  0.0000000000000000  0.8288012370000004  0.0649772224900005   F   F   F
  0.5000000000000000  0.3288012370000004  0.0649772224900005   F   F   F
  0.5000000000000000  0.8288012370000004  0.0649772224900005   F   F   F
  0.0000000000000000  0.4211987629999996  0.0649772224900005   F   F   F
  0.0000000000000000  0.9211987629999996  0.0649772224900005   F   F   F
  0.5000000000000000  0.4211987629999996  0.0649772224900005   F   F   F
  0.5000000000000000  0.9211987629999996  0.0649772224900005   F   F   F
  0.5000000000000000  0.6383952301165937  0.5418330154266708   T   T   T
  0.5000000000000000  0.8655653866421800  0.5418964543964928   T   T   T

Coverage 0.5 ML
  1.00000000000000
     7.6664519310000001    0.0000000000000000    0.0000000000000000
     0.0000000000000000   15.3329038620000002    0.0000000000000000
     0.0000000000000000    0.0000000000000000   24.3945007324000009
   Si   H
   64   20
Selective dynamics
Direct
  0.0000000000000000  0.1650190909834373  0.4958375084415420   T   T   T
  0.0000000000000000  0.6712214414604847  0.4841945621101842   T   T   T
```

```
0.5000000000000000  0.1883635882729676  0.4655860939282388  T T T
0.5000000000000000  0.6718484515338656  0.4843561923844414  T T T
0.0000000000000000  0.3116363967270369  0.4655857169282396  T T T
0.0000000000000000  0.8281514884661294  0.4843561923844414  T T T
0.5000000000000000  0.3349808940165664  0.4958375084415420  T T T
0.5000000000000000  0.8287784985395104  0.4841945621101842  T T T
0.2343257124120238  0.1315407130048367  0.4348889177342627  T T T
0.2503741156500194  0.6319750575291816  0.4341498463097261  T T T
0.7656742875879696  0.1315407130048367  0.4348889177342627  T T T
0.7496258843499752  0.6319750575291816  0.4341498463097261  T T T
0.2656742875879743  0.3684592869951624  0.4348889177342627  T T T
0.2496258843499791  0.8680249424708184  0.4341498463097261  T T T
0.7343257124120239  0.3684592869951624  0.4348889177342627  T T T
0.7503741156500248  0.8680249424708184  0.4341498463097261  T T T
0.2491373389958195  0.0001578440589922  0.3835692383303725  T T T
0.2508626610041826  0.4998421559410086  0.3835692383303725  T T T
0.7508626610041830  0.0001578440589922  0.3835692383303725  T T T
0.7491373389958170  0.4998421559410086  0.3835692383303725  T T T
0.2500000000000000  0.2500000000000000  0.3719306701640548  T T T
0.2500000000000000  0.7500000000000000  0.3728449591953378  T T T
0.7500000000000000  0.2500000000000000  0.3719306701640548  T T T
0.7500000000000000  0.7500000000000000  0.3728449591953378  T T T
0.0000000000000000 -0.0009652691135535  0.3259115096264308  T T T
0.0000000000000000  0.4990723819297934  0.3265864269672206  T T T
0.5000000000000000  0.0009276180702047  0.3265864269672206  T T T
0.5000000000000000  0.5009652691135599  0.3259115096264308  T T T
0.0000000000000000  0.2483105403028411  0.3176422596748155  T T T
0.0000000000000000  0.7495668692912012  0.3188459488346723  T T T
0.5000000000000000  0.2516894596971584  0.3176422596748155  T T T
0.5000000000000000  0.7504331307087988  0.3188459488346723  T T T
0.0000000000000000  0.1208628085531088  0.2662642086986366  T T T
0.0000000000000000  0.6217806477702584  0.2671490771930498  T T T
0.5000000000000000  0.1228363840137899  0.2668878293923095  T T T
0.5000000000000000  0.6230785473576828  0.2667301209094595  T T T
0.0000000000000000  0.3771636009862132  0.2668878293923095  T T T
0.0000000000000000  0.8769214526423172  0.2667301209094595  T T T
0.5000000000000000  0.3791371764468984  0.2662642086986366  T T T
0.5000000000000000  0.8782193522297416  0.2671490771930498  T T T
0.2503646594853965  0.1237902723033557  0.2111209337395403  T T T
0.2497264316911048  0.6238827145390345  0.2113137407407215  T T T
0.7496353405146032  0.1237902723033557  0.2111209337395403  T T T
0.7502735683088914  0.6238827145390345  0.2113137407407215  T T T
0.2496353405146023  0.3762097196966436  0.2111209337395403  T T T
0.2502735683088934  0.8761172854609655  0.2113137407407215  T T T
0.7503646594853968  0.3762097196966436  0.2111209337395403  T T T
0.7497264316911086  0.8761172854609655  0.2113137407407215  T T T
0.2500000000000000  0.0000000000000000  0.1555555609999999  F F F
0.2500000000000000  0.5000000000000000  0.1555555609999999  F F F
0.7500000000000000  0.0000000000000000  0.1555555609999999  F F F
0.7500000000000000  0.5000000000000000  0.1555555609999999  F F F
```

```
  0.2500000000000000  0.2500000000000000  0.1555555609999999  F  F  F
  0.2500000000000000  0.7500000000000000  0.1555555609999999  F  F  F
  0.7500000000000000  0.2500000000000000  0.1555555609999999  F  F  F
  0.7500000000000000  0.7500000000000000  0.1555555609999999  F  F  F
  0.0000000000000000  0.0000000000000000  0.1000000009999980  F  F  F
  0.0000000000000000  0.5000000000000000  0.1000000009999980  F  F  F
  0.5000000000000000  0.0000000000000000  0.1000000009999980  F  F  F
  0.5000000000000000  0.5000000000000000  0.1000000009999980  F  F  F
  0.0000000000000000  0.2500000000000000  0.1000000009999980  F  F  F
  0.0000000000000000  0.7500000000000000  0.1000000009999980  F  F  F
  0.5000000000000000  0.2500000000000000  0.1000000009999980  F  F  F
  0.5000000000000000  0.7500000000000000  0.1000000009999980  F  F  F
  0.0000000000000000  0.0788012370000004  0.0649772224900005  F  F  F
  0.0000000000000000  0.5788012370000004  0.0649772224900005  F  F  F
  0.5000000000000000  0.0788012370000004  0.0649772224900005  F  F  F
  0.5000000000000000  0.5788012370000004  0.0649772224900005  F  F  F
  0.0000000000000000  0.1711987679999964  0.0649772224900005  F  F  F
  0.0000000000000000  0.6711987679999964  0.0649772224900005  F  F  F
  0.5000000000000000  0.1711987679999964  0.0649772224900005  F  F  F
  0.5000000000000000  0.6711987679999964  0.0649772224900005  F  F  F
  0.0000000000000000  0.3288012370000004  0.0649772224900005  F  F  F
  0.0000000000000000  0.8288012370000004  0.0649772224900005  F  F  F
  0.5000000000000000  0.3288012370000004  0.0649772224900005  F  F  F
  0.5000000000000000  0.8288012370000004  0.0649772224900005  F  F  F
  0.0000000000000000  0.4211987629999996  0.0649772224900005  F  F  F
  0.0000000000000000  0.9211987629999996  0.0649772224900005  F  F  F
  0.5000000000000000  0.4211987629999996  0.0649772224900005  F  F  F
  0.5000000000000000  0.9211987629999996  0.0649772224900005  F  F  F
  0.0000000000000000  0.6360706303733108  0.5416732646921552  T  T  T
  0.5000000000000000  0.6370071324552682  0.5419049650343662  T  T  T
  0.0000000000000000  0.8629928675447318  0.5419049650343662  T  T  T
  0.5000000000000000  0.8639293696266892  0.5416732646921552  T  T  T

Coverage 0.75 ML
   1.00000000000000
     7.6664519310000001    0.0000000000000000    0.0000000000000000
     0.0000000000000000   15.3329038620000002    0.0000000000000000
     0.0000000000000000    0.0000000000000000   24.3945007324000009
   Si   H
   64   22
Selective dynamics
Direct
  0.0000000000000000  0.1761855468016103  0.4848477034885154  T  T  T
  0.0000000000000000  0.6709399801131666  0.4840636173774715  T  T  T
  0.5000000000000000  0.1947133585199044  0.4661869476922554  T  T  T
  0.5000000000000000  0.6712796425699763  0.4841728927481597  T  T  T
  0.0000000000000000  0.3327798190305383  0.4836179413251510  T  T  T
  0.0000000000000000  0.8278680320917825  0.4841904811990852  T  T  T
  0.5000000000000000  0.3397769363645051  0.4934129130757960  T  T  T
```

```
0.5000000000000000  0.8281999675791046  0.4841430182311897  T  T  T
0.2433922549535901  0.1315047068396269  0.4334951623723969  T  T  T
0.2502415870407844  0.6315244429715479  0.4339757473538943  T  T  T
0.7566077450464127  0.1315047068396269  0.4334951623723969  T  T  T
0.7497584129592180  0.6315244429715479  0.4339757473538943  T  T  T
0.2564604303432861  0.3674488385164693  0.4351027410262518  T  T  T
0.2498496460497347  0.8675269035421920  0.4340677177891369  T  T  T
0.7435395696567172  0.3674488385164693  0.4351027410262518  T  T  T
0.7501503539502717  0.8675269035421920  0.4340677177891369  T  T  T
0.2498159014803450 -0.0008039231761070  0.3830781319019887  T  T  T
0.2508302114852891  0.4990444225069662  0.3838022750904104  T  T  T
0.7501840985196545 -0.0008039231761070  0.3830781319019887  T  T  T
0.7491697885147172  0.4990444225069662  0.3838022750904104  T  T  T
0.2474232319090420  0.2520790234347795  0.3720837779558205  T  T  T
0.2500240004133084  0.7495924991939574  0.3727414125777521  T  T  T
0.7525767680909607  0.2520790234347795  0.3720837779558205  T  T  T
0.7499759995866917  0.7495924991939574  0.3727414125777521  T  T  T
0.0000000000000000 -0.0006246069021612  0.3257091723440366  T  T  T
0.0000000000000000  0.4990564512584992  0.3267157632940293  T  T  T
0.5000000000000000  0.0001746891351326  0.3259368934216278  T  T  T
0.5000000000000000  0.5001396706477793  0.3261935770565397  T  T  T
0.0000000000000000  0.2486555828262629  0.3170866863978548  T  T  T
0.0000000000000000  0.7493694295892643  0.3187796838104451  T  T  T
0.5000000000000000  0.2502953384496630  0.3191412035262055  T  T  T
0.5000000000000000  0.7497457675351956  0.3187566077542469  T  T  T
0.0000000000000000  0.1210588520743124  0.2660330988189400  T  T  T
0.0000000000000000  0.6215311541941431  0.2671613340959622  T  T  T
0.5000000000000000  0.1228003463066386  0.2670194107270265  T  T  T
0.5000000000000000  0.6222136654012635  0.2668047251339303  T  T  T
0.0000000000000000  0.3775370888823776  0.2665702806912824  T  T  T
0.0000000000000000  0.8767966080801111  0.2667616783492175  T  T  T
0.5000000000000000  0.3777253052965295  0.2669454220968149  T  T  T
0.5000000000000000  0.8773672155702089  0.2669198891316135  T  T  T
0.2506315425715114  0.1237630502803519  0.2110954546216720  T  T  T
0.2497588390176703  0.6238417097664414  0.2113503273863492  T  T  T
0.7493684574284828  0.1237630502803519  0.2110954546216720  T  T  T
0.7502411609823242  0.6238417097664414  0.2113503273863492  T  T  T
0.2502684818436895  0.3761167325671582  0.2112215590565506  T  T  T
0.2501110784919667  0.8760993456862717  0.2112465697275013  T  T  T
0.7497315181563071  0.3761167325671582  0.2112215590565506  T  T  T
0.7498889215080344  0.8760993456862717  0.2112465697275013  T  T  T
0.2500000000000000  0.0000000000000000  0.1555555609999999  F  F  F
0.2500000000000000  0.5000000000000000  0.1555555609999999  F  F  F
0.7500000000000000  0.0000000000000000  0.1555555609999999  F  F  F
0.7500000000000000  0.5000000000000000  0.1555555609999999  F  F  F
0.2500000000000000  0.2500000000000000  0.1555555609999999  F  F  F
0.2500000000000000  0.7500000000000000  0.1555555609999999  F  F  F
0.7500000000000000  0.2500000000000000  0.1555555609999999  F  F  F
0.7500000000000000  0.7500000000000000  0.1555555609999999  F  F  F
0.0000000000000000  0.0000000000000000  0.1000000009999980  F  F  F
```

```
0.000000000000000  0.500000000000000  0.100000000999980  F  F  F
0.500000000000000  0.000000000000000  0.100000000999980  F  F  F
0.500000000000000  0.500000000000000  0.100000000999980  F  F  F
0.000000000000000  0.250000000000000  0.100000000999980  F  F  F
0.000000000000000  0.750000000000000  0.100000000999980  F  F  F
0.500000000000000  0.250000000000000  0.100000000999980  F  F  F
0.500000000000000  0.750000000000000  0.100000000999980  F  F  F
0.000000000000000  0.078801237000004  0.064977222490005  F  F  F
0.000000000000000  0.578801237000004  0.064977222490005  F  F  F
0.500000000000000  0.078801237000004  0.064977222490005  F  F  F
0.500000000000000  0.578801237000004  0.064977222490005  F  F  F
0.000000000000000  0.171198767999964  0.064977222490005  F  F  F
0.000000000000000  0.671198767999964  0.064977222490005  F  F  F
0.500000000000000  0.171198767999964  0.064977222490005  F  F  F
0.500000000000000  0.671198767999964  0.064977222490005  F  F  F
0.000000000000000  0.328801237000004  0.064977222490005  F  F  F
0.000000000000000  0.828801237000004  0.064977222490005  F  F  F
0.500000000000000  0.328801237000004  0.064977222490005  F  F  F
0.500000000000000  0.828801237000004  0.064977222490005  F  F  F
0.000000000000000  0.421198762999996  0.064977222490005  F  F  F
0.000000000000000  0.921198762999996  0.064977222490005  F  F  F
0.500000000000000  0.421198762999996  0.064977222490005  F  F  F
0.500000000000000  0.921198762999996  0.064977222490005  F  F  F
0.000000000000000  0.137889637322420  0.541450786049203  T  T  T
0.000000000000000  0.636079804434516  0.541609518223803  T  T  T
0.500000000000000  0.636489124191257  0.541728445173104  T  T  T
0.000000000000000  0.365206288524185  0.541748694062889  T  T  T
0.000000000000000  0.863296758010178  0.541612581527392  T  T  T
0.500000000000000  0.863698227085057  0.541545744177916  T  T  T
```